\documentclass[aps,reprint,twocolumn,superscriptaddress,prd,nofootinbib]{revtex4-2}
\pdfoutput=1 
\usepackage[utf8]{inputenc}
\usepackage{url} 
\usepackage{float}
\usepackage{amsmath, amssymb}
\usepackage{comment}
\usepackage{array}
\usepackage[dvipsnames]{xcolor}
\usepackage{tcolorbox}
\usepackage{hyperref} 
\hypersetup{colorlinks=true, citecolor=blue, linkcolor=cyan}

\usepackage{graphicx} 
\usepackage{subcaption} 
\usepackage{cleveref} 
\usepackage[toc,page]{appendix} 

\definecolor{pinegreen}{rgb}{0.0, 0.47, 0.44}

\usepackage{natbib} 

\usepackage{tikz} 
\usetikzlibrary{shapes.geometric, arrows}
\tikzstyle{fkksflow} = [rectangle,rounded corners,text centered,fill = 
blue!30, text width = 7cm]
\tikzstyle{arrow} = [thick,->,>=stealth]

\Crefname{figure}{figure}{Figs.} 
\Crefname{equation}{Eq.}{Eqs.} 

\begin{document}

    \title{To bounce or not to bounce in generalized Proca theory and beyond}
     \author{Lara Bohnenblust}
    \affiliation{Department of Astrophysics, University of Zurich, Winterthurerstrasse 190, 8057 Zurich, Switzerland}

    \author{Serena Giardino}
    \email{serena.giardino@aei.mpg.de}
    \affiliation{Institute for Theoretical Physics, Universit{\"a}t Heidelberg, Philosophenweg 16, 69120 Heidelberg, Germany}
    \affiliation{Max Planck Institute for Gravitational Physics (Albert Einstein Institute), Am Mühlenberg 1, 14476 Potsdam, Germany}

    \author{Lavinia Heisenberg}
	\email{l.heisenberg@thphys.uni-heidelberg.de}
	\affiliation{Institute for Theoretical Physics, Universit{\"a}t Heidelberg, Philosophenweg 16, 69120 Heidelberg, Germany}

    \author{Nadine Nussbaumer}
    \email{nussbaumer\_n@thphys.uni-heidelberg.de}
    \affiliation{Institute for Theoretical Physics, Universit{\"a}t Heidelberg, Philosophenweg 16, 69120 Heidelberg, Germany}

    \begin{abstract}
        It is notoriously difficult to construct a stable non-singular bouncing cosmology that avoids all possible instabilities throughout the entire evolution of the universe. In this work, we explore whether a non-singular bounce driven by a specific class of modifications of General Relativity, the vector-tensor generalized Proca theories, can be constructed without encountering any pathologies in linear perturbation theory. We find that such models unavoidably lead either to strong coupling in the tensor or the scalar sector, or instabilities in the matter sector during the bouncing phase. As our analysis is performed in a gauge-independent way, this result can be cast in the form of a no-go theorem for non-singular bounces with generalized Proca. In contrast to the no-go theorem found for Horndeski theories, however, it cannot be evaded by considering beyond generalized Proca theory. At the core of our result lies the non-dynamical nature of the temporal component of the vector field, which renders it an ill-suited mediator for a bouncing solution.
    \end{abstract}

    \maketitle

    \section{Introduction}
        Inflation is arguably the most developed and successful paradigm to describe the evolution of the early universe. Not only does this phase of almost exponential expansion address a number of problems faced by standard Big Bang cosmology, such as the horizon and the flatness problems, but it also provides a causal mechanism to generate primordial fluctuations that are stretched to super-horizon scales and constitute the seeds of all structure in the universe \cite{Mukhanov:2005sc}. This leads to viable predictions for the power spectrum of the cosmic microwave background (CMB) radiation and its tilt that are in accordance with observations \cite{Planck:2018jri, Planck:2018vyg}. Nonetheless, the theory of inflation suffers from several shortcomings \cite{Brandenberger:1999sw}: on the one hand, inflationary cosmologies necessarily contain initial singularities \cite{Borde:1996pt} and are thus geodesically past-incomplete, which leaves the question about the beginning of the universe unanswered. On the other hand, the duration of inflation is extremely fine-tuned in standard inflationary models driven by a scalar field \cite{Martin:2000xs}. If the exponential expansion takes only slightly longer than needed in order to solve the problems of Big Bang cosmology, the wavelengths of modes corresponding to cosmological scales today must have originated in the trans-Planckian regime, where the underlying effective field theory description is known to break down and General Relativity (GR) needs to be replaced by the yet unknown theory of quantum gravity (this is known as \textit{trans-Planckian problem}). Additionally, inflation, in its modern reading, is understood as a broad paradigm that encompasses a variety of specific models: the textbook version of a single, minimally-coupled scalar field with a simple quadratic potential has been made obsolete not only by more complicated potentials (often \textit{a posteriori} including features that match observational data better), but also by models containing more than one field or more than one inflationary phase. This substantial flexibility of the inflationary paradigm constitutes a problem: in the past four decades since its inception, inflationary models leading to not only different, but also contradictory cosmological predictions have been devised and then fell into oblivion when new observations ruled them out. Moreover, most models of inflation lead to eternal inflation, which generates a multiverse in which predictability is lost, in the absence of a suitable probability measure. Despite the popularity of anthropic arguments related to string theory in this context, this line of research enters a risky realm where falsifiability is at stake. These issues have been pointed out and discussed, for example, in \cite{Ijjas:2013vea, Ijjas:2014nta, Guth:2013sya, Chowdhury:2019otk, Guth:2007ng}. Furthermore, how robust inflation is with respect to different initial conditions represents a concern. This is currently being investigated with the cutting-edge tools of numerical relativity, which allow to consider wildly varying initial conditions beyond the linear regime \cite{East:2015ggf, Clough:2016ymm, Clough:2017efm, Corman:2022alv, Elley:2024alx, Aurrekoetxea:2024mdy}.

        These shortcomings of the inflationary paradigm motivate the exploration of alternatives. Even if inflation turned out to be the best paradigm for the early universe, addressing the challenges posed by alternative models give it the chance to grow more solid. Furthermore, many of the inflationary predictions are not unique to inflation, and observational "smoking guns" of inflation have not yet been found (and are indeed quite difficult to single out \cite{Brandenberger:2011eq}). For example, a decade before the first inflationary models were suggested, it had already been realised \cite{Sunyaev:1970bma, Peebles:1970ag} that an approximately scale-invariant spectrum of adiabatic perturbations on scales larger than the Hubble horizon at the time of matter-radiation equality leads to precisely the power spectrum that has been subsequently observed \cite{Planck:2018jri}. Inflation is historically the \textit{first} mechanism that has been able to generate from first principles a primordial spectrum of cosmological fluctuations that seeds cosmic structure, but it is not the \textit{only} one \cite{Brandenberger:2003vk}. Thus, it is important to work out the predictions of other falsifiable paradigms for early universe cosmology, which might allow to distinguish inflation from alternatives. As the next generation of CMB observatories is under way (e.g. \cite{CMB-S4:2022ght}), upcoming data of unprecedented precision will bear the fingerprints of such early universe processes and help shed light on the correct paradigm.
        
        An intriguing alternative to inflation, which evades the initial singularity as well as the trans-Planckian problem, is provided by bouncing cosmologies (see \cite{Brandenberger:2016vhg, Battefeld:2014uga} for reviews), in which the Big Bang is replaced by a bounce. Bouncing cosmologies have a long history, especially in the context of cyclic universe models that expand and contract periodically. They were first explored by ancient philosophers, and then by Einstein \cite{Einstein1931} and Tolman \cite{Tolman}, among others, witnessing a more recent surge of interest with the development of the ekpyrotic scenario \cite{Khoury:2001wf}. Bouncing models where the singularity is avoided by virtue of some quantum gravity proposal have been explored, for example, in the context of Loop Quantum Cosmology \cite{Ashtekar:2011ni, Cai:2014zga}. However, in the absence of a fully-fledged consistent theory of quantum gravity, non-singular bounces where the scale factor never crosses zero and the evolution can be treated purely classically, are an attractive possibility.
        
        During a classical non-singular bounce, the universe transitions from a contracting to an expanding phase, always keeping a finite size and energy density well below the Planck scale. At the moment of the bounce $t_B$, the scale factor $a(t_B)$ hence assumes a finite positive value, while the Hubble parameter $H(t)=\Dot{a}(t)/a(t)$ describing the expansion rate of the universe crosses zero at $t_B$. Therefore, the initial singularity present in inflationary scenarios can be avoided. Furthermore, cosmological fluctuations originate in the sub-horizon regime well above the Planck scale and become super-horizon as the universe contracts, i.e.\ they proceed analogously to inflationary fluctuations. In that sense, bouncing cosmologies also allow for mechanisms to generate initial perturbations for structure formation, with the exact process being determined by the bouncing model at hand. The length scale of fluctuations corresponding to cosmological scales today remain several orders of magnitude larger than the Planck scale and thus bouncing cosmologies do not suffer from the trans-Planckian problem \cite{Brandenberger:2016vhg}. The evolution of the universe after the bounce ends is precisely the same as in standard inflationary cosmology after the end of inflation. \\
        When attempting to construct a non-singular bounce in the context of GR with matter coupled minimally to gravity, there are many obstacles to tackle, already at the background level. The most relevant one is set by the singularity theorems of Penrose and Hawking \cite{Hawking:1973uf}, which state that a singularity at the beginning of the universe is unavoidable if gravity is described by GR in the presence of standard matter obeying the Null Energy Condition (NEC). The NEC for a perfect fluid reads $\rho_M+P_M\geq 0$, with $\rho_M$ and $P_M$ the energy density and pressure of the fluid, respectively. A bouncing phase requires $\Dot{H}>0$ to accommodate the switch from contraction ($H<0$) to expansion ($H>0$), but this is not achievable in GR with standard matter, since the background equations of GR in a  Friedmann-Lema\^itre-Robertson-Walker (FLRW) setting imply
        \begin{equation}
            \Dot{H}=-\frac{1}{2M_\mathrm{Pl}^2}(\rho_M+P_M)
        \end{equation}
        (where $M_{\rm Pl}$ is the Planck mass) which directly shows that $\Dot{H}<0$ in general if the NEC is valid. Therefore, achieving a bouncing phase requires the violation of the NEC: this can be done either with ghost matter ($\rho_M+P_M\leq 0$) \cite{Cai:2007qw, Lin:2010pf}, which generally introduces instabilities, or by modifying gravity by going beyond GR, which is the route that we are interested in here. The simplest possibility to modify gravity is introducing new degrees of freedom (DOF) to GR \cite{Heisenberg:2018vsk}. However, in order to obtain a consistent theory, it has to be ensured that the new DOF do not introduce pathologies such as ghost, gradient and tachyonic instabilities or strong coupling, once linear perturbations are introduced (see Appendix A of \cite{Heisenberg:2018vsk} for details). Furthermore, it has been shown in \cite{Dubovsky:2005xd} that NEC-violation itself commonly leads to ghosts and other generic instabilities already at the classical level, which clearly demonstrates that bounces emerging from modified gravity theories have to be carefully analysed regarding their stability over the entire cosmological evolution (and not only near the bounce). 
        
        A solid and well-studied candidate for stable ghost-free modified gravity is given by \textit{Horndeski theory} \cite{Horndeski:1974wa, Horndeski:2024sjk}, the most general class of scalar-tensor theories with derivative self-interactions keeping second-order equations of motion (EOM). So far, extensive research has gone into studying bounces in several sub-classes of Horndeski theory, e.g. \cite{Osipov:2013ssa, Qiu:2011cy,  Easson:2011zy, Cai:2012va, Battarra:2014tga, Mironov:2019haz, Mironov:2024pjt}. However, there seems to always be some obstruction to constructing \textit{complete} non-singular cosmological scenarios in Horndeski theory, i.e., where the evolution of the universe can be followed from $t\rightarrow -\infty$ to $t \rightarrow +\infty$ without encountering any instability throughout. The exploration of these issues culminated in the formulation of a no-go theorem for such models \cite{Libanov:2016kfc,Kobayashi:2016xpl}, that can be circumvented by considering \textit{beyond Horndeski theory} \cite{Gleyzes:2014dya,Mironov:2019haz,Cai:2017dyi,Kolevatov:2017voe}, an extension which remains ghost-free but allows for stable higher-order equations of motion (see \cite{Ijjas:2016vtq} for a different perspective on Horndeski bounces).
        
        Another intriguing possibility of constructing stable non-singular bounces is given by Horndeski's vector pendant, \textit{generalized Proca theory} \cite{Heisenberg:2014rta} in curved spacetime, namely the most general class of vector-tensor theories with second-order EOM which maintains three healthy propagating degrees of freedom.
        Intriguingly, studies of the broad class of scalar-vector-tensor (SVT) theories (encompassing both Horndeski and generalized Proca as specific cases) found that scalar-tensor theories generically incur into strong coupling problems, while vector-tensor theories do not and thus present an advantage \cite{Heisenberg:2018wye}, as we discuss in more detail in Section \ref{pert}. The analysis in \cite{Heisenberg:2018wye} does not rely on a specific gauge choice, which can be a deal-breaker when analysing perturbative stability, since perturbation variables in different gauges might be ill-suited for particular applications such as bounces. 
        So far, conditions to avoid ghost and gradient instabilities have been computed for generalized Proca theory in \cite{Heisenberg:2018wye,DeFelice:2016uil,DeFelice:2016yws}, but whether a stable bounce can be achieved has remained an open question, which we address in the present work. Specifically, we investigate whether a linearly stable non-singular bouncing solution driven by generalized Proca can be achieved.
        
        In the present work, after reviewing generalized Proca in Section \ref{proca} and their cosmological perturbations in Section \ref{pert}, we present a no-go theorem for non-singular bounces in such theories in Section \ref{nogo}, showing that instabilities of some sort will always be encountered either in the matter, tensor or scalar sector. We also explain why the theorem cannot be evaded by employing \textit{beyond generalized Proca theory} \cite{Heisenberg:2016eld}, the beyond Horndeski equivalent for vector fields, which keeps three dynamical DOF with stable higher-order EOM. 

    \section{Generalized Proca theory and background dynamics}
    \label{proca}
        The development of generalized Proca theories was inspired by the realisation that it is not possible to construct non-trivial self-interactions for a $U(1)$ gauge-invariant, massless vector field, while naively extending results from scalar-tensor theories \cite{Deffayet:2013tca}. However, this feat could be achieved instead for a \textit{massive} vector field, which breaks gauge invariance, propagates three healthy DOF (the third being the longitudinal polarization, in addition to the usual transverse ones) and maintains second-order EOM, avoiding the Ostrogradski instability due to the structure of the self-interactions in $\mathcal{L}_{i>2}$.
        These are the so-called generalized Proca theories, first found in \cite{Heisenberg:2014rta}, whose action (in curved spacetime) reads
        \begin{equation}\label{action}
            S=\int d^4x \sqrt{-g}\left(\mathcal{L}+\mathcal{L}_M\right),\hspace{20pt}\mathcal{L}=\sum_{i=2}^6\mathcal{L}_i,
        \end{equation}
        where $g$ denotes the metric determinant, $\mathcal{L}_M$ is a matter action and the Lagrangians $\mathcal{L}_i$ are given by 
        \begin{align}
        \label{L2}
            &\mathcal{L}_2=G_2(X,Y,F)\\
            &\mathcal{L}_3=G_3(X)\nabla_\mu A^\mu\\
            &\mathcal{L}_4=G_4(X)R+G_{4,X}(X)[(\nabla_\mu A^\mu)^2-\nabla_\rho A^\sigma\nabla_\sigma A^\rho]
            \end{align}
            \begin{align}
            &\mathcal{L}_5=G_5(X)G_{\mu\nu}\nabla^\mu A^\nu-\frac{1}{6}G_{5,X}(X)[(\nabla_\mu A^\mu)^3\notag\\ 
            &\hspace{20pt}+2\nabla_\rho A_\sigma\nabla^\gamma A^\rho\nabla^\sigma A_\gamma-3(\nabla_\mu A^\mu)\nabla_\rho A_\sigma \nabla^\sigma A^\rho]\notag\\ &\hspace{20pt}-g_5(X)\Tilde{F}^{\alpha\mu}\Tilde{F}^\beta_\mu\nabla_\alpha A_\beta\\
            &\mathcal{L}_6=G_6(X)L^{\mu\nu\alpha\beta}\nabla_\mu A_\nu\nabla_\alpha A_\beta\notag\\&\hspace{20pt}+\frac{1}{2}G_{6,X}(X)\Tilde{F}^{\alpha\beta}\Tilde{F}^{\mu\nu}\nabla_{\alpha}A_\mu\nabla_\beta A_\nu,
            \label{L6}
        \end{align}
        where $\nabla_\mu$ is the covariant derivative and $A_\mu$ the generalized Proca field with associated field strength tensor $F_{\mu\nu}=\nabla_\mu A_\nu-\nabla_\nu A_\mu$. The vector field is coupled non-minimally to gravity via the coupling functions $G_4$, $G_5$ and $G_6$. While the function $G_2$ depends on the scalar quantities 
        \begin{align}
        \label{X}
            &X=-\frac{1}{2}A_\mu A^\mu,\\
            &F=-\frac{1}{4}F_{\mu\nu}F^{\mu\nu},\\
            &Y=A^\mu A^\nu F_\mu^\alpha F_{\nu\alpha},
        \end{align}
        the other coupling functions $G_{i>2}$ and $g_5$ are functions of $X$ only. Partial derivatives are denoted by $G_{i,X}=\partial G_i/\partial X$. Furthermore, we introduced the dual field strength tensor and the double dual Riemann tensor
        \begin{align}
            \Tilde{F}^{\mu\nu}=\frac{1}{2}\epsilon^{\mu\nu\alpha\beta}F_{\alpha\beta},\hspace{20pt}L^{\mu\nu\rho\sigma}=\frac{1}{4}\epsilon^{\alpha\beta\gamma\delta}R_{\rho\sigma\gamma\delta},
        \end{align}
        with $\epsilon^{\mu\nu\rho\sigma}$ the Levi-Civita tensor and $R_{\rho\sigma\gamma\delta}$ the Riemann tensor.
        The original Proca Lagrangian can be recovered by setting $G_2(X)=m^2 X$ and all derivative self-interactions $G_{i>2}=0$. Therefore, $G_2$ can be viewed as a generalized potential term including all possible kinetic and interaction contributions assembled from the vector field $A_\mu$, its field strength tensor $F_{\mu\nu}$ and its dual $\Tilde{F}^{\mu\nu}$. Upon taking the limit $A_{\mu}\rightarrow \nabla_{\mu}\pi$ (where $\pi$ denotes a scalar field), one recovers the Horndeski scalar interactions.\\

        The generalized Proca theory given in \eqref{L2}-\eqref{L6} can be extended to \textit{beyond generalized Proca theory} by allowing for higher-order EOM that do not suffer from the Ostrogradsky instability (similarly to beyond Horndeski theories \cite{Gleyzes:2014dya}) due to the existence of a specific constraint in the vector case that ensures the right number of propagating DOF. The authors of \cite{Heisenberg:2016eld} have found the extra terms 
        \begin{align}
        \label{L4N}
            &\mathcal{L}_4^N=f_4(X)\epsilon^{\mu\nu\rho\sigma}\epsilon^{\alpha\beta\delta}_\sigma A_\alpha A_\mu\nabla_\beta A_\nu\nabla_\delta A_\rho,\\
            &\mathcal{L}_5^N=f_5(X)\epsilon^{\mu\nu\rho\sigma}\epsilon^{\alpha\beta\delta\kappa} A_\alpha A_\mu\nabla_\beta A_\nu\nabla_\delta A_\rho\nabla_\sigma A_\kappa,\\
            &\Tilde{\mathcal{L}}_5^N=\Tilde{f}_5(X)\epsilon^{\mu\nu\rho\sigma}\epsilon^{\alpha\beta\delta\kappa} A_\alpha A_\mu\nabla_\beta A_\delta\nabla_\nu A_\rho\nabla_\sigma A_\kappa,\label{L5barN}\\
            &\mathcal{L}_6^N=\Tilde{f}_6(X)\epsilon^{\mu\nu\rho\sigma}\epsilon^{\alpha\beta\delta\kappa} \nabla_\alpha A_\beta \nabla_\mu A_\nu\nabla_\delta A_\rho\nabla_\kappa A_\sigma,
            \label{L6N}
        \end{align}
        where $f_4, f_5, \Tilde{f}_5$ and $\Tilde{f}_6$ are arbitrary scalar functions of $X$. Notice that $\Tilde{\mathcal{L}}_5^N$ and $\mathcal{L}_6^N$ comprise the intrinsic vector mode contributions. The Lagrangians \eqref{L4N}-\eqref{L6N} can then be added to \eqref{action} to yield the full beyond generalized Proca theory on a curved background
        \begin{equation}
        \label{actionbeyond}
            S=\int d^4x\sqrt{-g}(\mathcal{L}+\mathcal{L}^N+\mathcal{L}_M),
        \end{equation}
        with
        \begin{equation}
        \label{Lbeyond}
            \mathcal{L}^N=\mathcal{L}_4^N+\mathcal{L}_5^N+\Tilde{\mathcal{L}}_5^N+\mathcal{L}_6^N.
        \end{equation}
        Higher order EOM in beyond generalized Proca theory arise directly from the terms \eqref{L4N}-\eqref{L5barN}. When adding $\mathcal{L}_6^N$ to $\mathcal{L}_6$, the relative coefficient in front of the second term of \eqref{L6} becomes detuned. Given that the relative coefficients in generalized Proca theory \eqref{L2}-\eqref{L6} are chosen such that second order EOM can be maintained, the detuning in $\mathcal{L}_6$ in beyond generalized Proca theory directly leads to higher-order EOM.

        \subsection{Background}
            On a flat FLRW background with metric signature $(-,+,+,+)$, the vector field takes the simple form
            \begin{equation}\label{AFLRW}
                A^\mu=(-\bar{A}_0(t),0,0,0),
            \end{equation}
            as isotropy forbids it to point in any direction. Additionally, owing to spatial homogeneity, the temporal component can only be a function of time $t$. Given the vector field \eqref{AFLRW} and the FRLW background, we readily find $X= \bar{A}_0^2/2$ as well as $F_{\mu\nu}=\Tilde{F}^{\mu\nu}=L^{\mu\nu\rho\sigma}=0$.\\
            Generalized Proca theories have intriguing applications in cosmology and exhibit a rich phenomenology \cite{DeFelice:2016yws, Heisenberg:2020xak}. Specifically, the temporal component of the vector field $\bar{A}_0(t)$ can drive the late-time cosmic acceleration because of its derivative interactions (compatibly with observational constraints) and there exist de Sitter solutions and a de Sitter fixed point that is always stable, as investigated in detail in \cite{DeFelice:2016yws}.\\
            
            We model the matter component $\mathcal{L}_M$ by a perfect fluid minimally coupled to gravity. Given the energy density $\rho_M$ and isotropic pressure $P_M$ of the fluid, the continuity equation
            \begin{equation}
            \label{conteq}
                \Dot{\rho}_M+3H(\rho_M+P_M)=0
            \end{equation}
            holds.
            Varying the action \eqref{action} with respect to the metric tensor $g_{\mu\nu}$, we find the background equations of motion \cite{DeFelice:2016yws}
            \begin{align}\label{rho}
                &G_2-G_{2,X}\bar{A}_0^2+3G_{3,X}H \bar{A}_0^3+6G_4 H^2\notag\\&\hspace{12pt}-6(2G_{4,X}+G_{4,XX}\bar{A}_0^2)H^2\bar{A}_0^2-G_{5,XX}H^3\bar{A}_0^5\notag\\&\hspace{12pt}-5G_{5,X}H^2\bar{A}_0^3=\rho_M,\\
                &G_2+\dot{\bar{A}}_0\bar{A}_0^2G_{3,X}+2G_4(3H^2+2\Dot{H})\notag\\&\hspace{12pt}-2G_{4,X}\bar{A}_0(3H^2\bar{A}_0+2H\dot{\bar{A}}_0+2\Dot{H}\bar{A}_0)\notag\\&\hspace{12pt}-4G_{4,XX}H\dot{\bar{A}}_0\bar{A}_0^3-G_{5,XX}H^2\dot{\bar{A}}_0\bar{A}_0^4\notag\\&\hspace{12pt}-G_{5,X}H\bar{A}_0^2(2\Dot{H}\bar{A}_0+2H^2\bar{A}_0+3H\dot{\bar{A}}_0)=-P_M.\label{p}
            \end{align}
            It is straightforward to check that \eqref{rho}-\eqref{p} fulfil the continuity equation \eqref{conteq}. Notice that $\mathcal{L}_6$ does not influence the background EOM since it corresponds to the intrinsic vector contribution, whereas on a FLRW background, the vector field is only allowed to have a scalar temporal component \eqref{AFLRW}.
            Variation of \eqref{action} with respect to $\bar{A}_0$ leads to 
            \begin{align}\label{A0eq}
                \mathcal{E}_{\bar{A}_0}&= \bar{A}_0(G_{2,X}-3G_{3,X}H\bar{A}_0+6H^2(G_{4,X}+\bar{A}_0^2G_{4,XX})\notag\\&\hspace{10pt}+\bar{A}_0H^3(3G_{5,X}+\bar{A}_0^2G_{5,XX})=0.
            \end{align}
            Ignoring the trivial solution branch $\bar{A}_0=0$, the equation of motion \eqref{A0eq} clearly demonstrates that the field $\bar{A}_0$ is non-dynamical as the equation is algebraic; its behaviour is simply determined by the cosmological background $H(t)$. One can hardly understate the importance of this fact, especially as this non-dynamical character of the vector field is so different with respect to Horndeski theory, which describes a fully dynamical scalar degree of freedom (see section \ref{nogo} for a discussion on this aspect).

            The background dynamics of beyond generalized Proca theory \eqref{actionbeyond}-\eqref{Lbeyond} was previously examined in \cite{Heisenberg:2016eld}, which found that this theory is indistinguishable from generalized Proca theory at the background level. This boils down to the fact that, after cleverly encoding the additional couplings of $\mathcal{L}_4^N, \mathcal{L}_5^N$ in the two functions $B_4(X)$ and $B_5(X)$, the latter do not appear in the background EOM at all. Note that $\Tilde{\mathcal{L}}_5^N$ and $\mathcal{L}_6^N$ generally do not enter the dynamics of the background as they only correspond to intrinsic vector modes.

 
    \section{Cosmological perturbations}
    \label{pert}
        The most relevant roadblock faced in constructing a viable non-singular bouncing solution is ensuring its linear stability around the bounce. This is a very challenging task that halted research efforts on bouncing models for quite a long time. Anisotropies can grow and disrupt the bounce, and the contracting phase is especially delicate (see, e.g. \cite{Battefeld:2014uga} for an overview of these issues). Nevertheless, studying perturbations at the linear level is simply the first stage (which would be enough for a proof of principle that a stable bouncing solution could exist), but a more complete approach would also involve testing stability at the non-linear level, beyond small perturbations around the FLRW metric. 
        In the present work, we explore whether a linearly stable non-singular bouncing solution driven by generalized Proca exists. At the level of the action, this corresponds to performing an expansion up to second order in perturbations, which was previously done by \cite{Heisenberg:2018wye,DeFelice:2016yws} for the theory at hand, described by \eqref{action}. In this section, we thus briefly recap the key steps from \cite{Heisenberg:2018wye,DeFelice:2016yws} and adopt the same notation. \\
        Owing to the $SO(3)$ symmetry in the spatial part of the FLRW metric, we can perform an SVT decomposition of all relevant quantities. Up to second order, the scalar, vector and tensor parts do not mix in the expanded action and can hence be studied separately.
        For the gravitational sector, we consider the SVT-decomposed FLRW line element
        \begin{align}
        \label{ds}
            ds^2=&-(1+2\alpha)dt^2+2(V_i+\partial_i\chi)dtdx^i\notag\\&+a(t)^2[(1+2\zeta)\delta_{ij}+2\partial_i \partial_j E+h_{ij}]dx^idx^j,
        \end{align}
        where $h_{ij}$ refer to tensor perturbations, $V_i$ denote vector modes and $\alpha$, $\chi, \zeta$ and $E$ are scalar perturbations. While both the tensor and vector perturbations are transverse, i.e.\ satisfy $\partial^i h_{ij}=0$ and $\partial^iV_i=0$, the tensor modes are additionally traceless $h^i_i=0$. The action above is written in a \textit{gauge-ready formulation}, meaning it does not involve any gauge choice. It contains several non-dynamical fields that need to be integrated out before the dynamical degrees of freedom can be identified (see Section \ref{scalarperts}).
        These conditions leave us with two propagating tensor and two propagating vector DOF, in addition to only one dynamical scalar mode in the gravity sector (an additional propagating scalar is present in the matter sector).

        The generalized Proca field $A^\mu$ can be SVT-decomposed as 
        \begin{align}
            &A^0=-\bar{A}_0(t)+\delta A,\\
            &A^i=\frac{1}{a^2}\delta^{ij}(\partial_j\chi_V+\delta A_j),
        \end{align}
        exhibiting two scalar perturbations $\delta A$, $\chi_V$ as well as a vector perturbation $\delta A_j$ obeying the transversality constraint $\partial^j\delta A_j=0$.
        
        For the matter sector, a perfect fluid at the action level is conveniently described by the Schutz-Sorkin action \cite{Schutz:1977df,DeFelice:2016yws}
        \begin{equation}
        \label{schutzsorkin}
            S_M=-\int d^4x[\sqrt{-g}\rho_M(n)+J^\mu(\partial_\mu\ell+\mathcal{A}_1\partial_\mu\mathcal{B}_1+\mathcal{A}_2\partial_\mu\mathcal{B}_2],
        \end{equation}
        where $\ell$ is a scalar, $J^\mu$ a vector field, and the $\mathcal{A}_i$'s and $\mathcal{B}_i$'s are scalars describing transverse vector perturbations. The energy density $\rho_M$ is a function of the fluid's number density 
        \begin{equation}
            n=\sqrt{\frac{J^\mu J^\nu g_{\mu\nu}}{g}},
        \end{equation}
        which is directly linked to the vector field $J^\mu$. On the FLRW background, we can decompose the temporal and spatial parts of $J^\mu$ as
        \begin{align}
            &J^0=\mathcal{N}_0+\delta J,\\
            &J^i=\frac{1}{a^2}\delta^{ij}(\partial_j\delta j+W_j),
        \end{align}
        with corresponding scalar perturbations $\delta J$, $\delta j$ and a vector mode $W_i$ satisfying the transverse condition $\partial^i W_i=0$. $\mathcal{N}_0$ denotes the total fluid number and is related to the number density via $n_0=\mathcal{N}_0/a^3$. The scalar quantity $\ell$ can be decomposed as
        \begin{equation}
            \ell=-\int^t\rho_{M,n}dt'-\rho_{M,n}v,
        \end{equation}
        where $v$ stands for the velocity potential and is first-order in perturbations, while $\rho_{M, n} \equiv \partial \rho_M / \partial n$.
        The matter sound speed is
        \begin{equation}
        \label{cM}
            c_M^2\equiv \frac{\dot{P}_M}{\dot{\rho}_M}=\frac{n_0 \rho_{M, n n}}{\rho_{M, n}},
        \end{equation}
        where we used the fact that the background pressure can be expressed as $P_M=n_0\rho_{M,n}-\rho_M$ (for more details, see \cite{Heisenberg:2018wye}).
    
        \subsection{Tensor perturbations}
            Due to isotropy, we can consider each sector separately. We start with the least involved sector, namely the tensor modes $h_{ij}$ on top of the FLRW background in the line element \eqref{ds}. The transverse-traceless property of $h_{ij}$ allows us to choose the non-zero components as $h_{11}=h_1(t,z)$, $h_{22}=-h_1(t,z)$ and $h_{12}=h_{21}=h_2(t,z)$, where the functions $h_1$, $h_2$ represent the two propagating tensor degrees of freedom. For simplicity, we assumed that the perturbations have only the $z$-dependence. Expanding the action \eqref{action} up to second order in tensor perturbations, we obtain
            \begin{equation}
                S_T^{(2)}=\int dt\,d^3x\sum_{i=1}^2 q_T\left[\Dot{h}_i^2-\frac{c_T^2}{a^2}(\partial h_i)^2\right],
            \end{equation}
            with the stability coefficients \cite{DeFelice:2016yws, Heisenberg:2018wye}
            \begin{align}
                \label{qT}
                &q_T=2G_4-2\bar{A}_0^2G_{4,X}-\bar{A}_0^3HG_{5,X},\\[10pt]
                &c_T^2=\frac{2G_4-\bar{A}_0^2\dot{\bar{A}}_0G_{5,X}}{q_T}\equiv\frac{\mathcal{F}_T}{q_T}.\label{cT}
            \end{align}
            Equation \eqref{cT} is the squared propagation speed of the tensor modes. Avoiding ghost instability amounts to ensuring $q_T>0$, while evading the Laplacian instability requires $c_T^2\geq0$.

        \subsection{Vector perturbations}
            Next, we examine the FLRW line element \eqref{ds} solely perturbed by the vector field $V_i$. We let the vector perturbations $V_i=V_i(t,z)$ depend on $t, z$ only, in order to simplify the computation slightly but this does not introduce any fundamental restriction. When computing the quadratic action, it is useful to introduce the combination \cite{DeFelice:2016uil}
            \begin{equation}
                Z_i=\delta A_i-\bar{A}_0(t)V_i,
            \end{equation}
            so that $A_i=Z_i$ for vector perturbations only. Furthermore, we perform the rescaling 
            \begin{equation}
                \Tilde{Z}_i=\frac{Z_i}{a}.
            \end{equation}
            After integrating out all non-dynamical fields (such as e.g. the fluid vector perturbations) and taking the small-scale limit (or considering wavelengths $k^2\gg a^2 H
            $) as in \cite{Heisenberg:2018wye},
            we obtain the final second-order vector action \cite{DeFelice:2016uil}
            \begin{equation}\label{actionV}
                S_V^{(2)}\simeq\int dt\,d^3x\frac{a^3}{2}q_V\left[\Dot{\Tilde{Z}}_i^2-\frac{c_V^2}{a^2}(\partial_j\Tilde{Z}_i)^2\right],
            \end{equation}
            where clearly only two modes in $Z_i=(Z_1(t,z),Z_2(t,z),0)$ are propagating. Notice that we have dropped the effective mass term $m_V^2 Z_i^2$ in \eqref{actionV} as it becomes negligible compared to the gradient term in the small-scale limit for $c_V^2\neq 0$ (see also \cite{DeFelice:2016yws}). The stability coefficients are \cite{DeFelice:2016uil}
            \begin{align}
                &q_V=G_{2,F}+2\bar{A}_0^2G_{2,Y}-4\bar{A}_0Hg_5\notag\\&\hspace{22pt}+2H^2(G_6+\bar{A}_0^2G_{6,X})
            \end{align}
            and
            \begin{align}
                &c_V^2\equiv \frac{\mathcal{F}_V}{q_V}=1+\frac{\bar{A}_0^2(2G_{4,X}+\bar{A}_0HG_{5,X})^2}{2q_Tq_V}\\&\hspace{21pt}+\frac{2[\Dot{H}G_6-\bar{A}_0^2G_{2,Y}-(\bar{A}_0H-\dot{\bar{A}}_0)(\bar{A}_0HG_{6,X}-g_5)]}{q_V}\notag,
            \end{align}
            The absence of of ghosts is ensured by $q_V>0$ and the Laplacian instability at small scales is avoided for a positive vector speed of sound $c_V^2\geq0$. 
   
        \subsection{Scalar perturbations}
        \label{scalarperts}
            The quadratic action for scalar perturbations, which needs a more involved treatment, was provided in \cite{Heisenberg:2018wye} for SVT theories in a gauge-ready form. We will review the procedure here and adapt it to non-singular bouncing cosmologies in generalized Proca. We first perform the redefinition
            \begin{equation}
                \psi\equiv\chi_V-\bar{A}_0(t)\chi,
            \end{equation}
            such that the scalar part of the spatial Proca field perturbation becomes $A_i=\partial_i\psi$. Then, the final scalar-mode action at second order reads 
            \begin{equation}
             \label{actionS}
                S_S^{(2)}=\int dt\,d^3x\,(\mathcal{L}^{\mathrm{flat}}+\mathcal{L}_{\zeta}+\mathcal{L}_E),
            \end{equation}
            where
            \begin{widetext}
           \begin{align}
                \label{Lflat}
               \mathcal{L}^\mathrm{flat}=a^3&\Bigg[\left(\frac{w_2}{\bar{A}_0}(\partial_i\alpha)-w_1(\partial_i\delta A)\right)\frac{\partial^i\chi}{a^2}-w_3\frac{(\partial\alpha)^2}{a^2}+w_4\alpha^2+w_3\frac{(\partial_i\delta A)(\partial^i\alpha)}{a^2 \bar{A}_0}+w_3\left(\partial_i\alpha-\frac{1}{2\bar{A}_0}(\partial_i\delta A)\right)\frac{\partial_i\dot{\psi}}{a^2\bar{A}_0}\notag\\&+w_8\frac{\delta A\alpha}{\bar{A}_0}+\left(w_6(\partial_i\alpha)+\frac{w_2-\bar{A}_0 w_6}{2\bar{A}_0^2}(\partial_i\delta A)\right)\frac{\partial^i\psi}{a^2}-w_3\frac{(\partial\delta A)^2}{4a^2\bar{A}_0^2}+w_5\frac{\delta A^2}{\bar{A}_0^2}-w_3\frac{(\partial\dot{\psi})^2}{4a^2\bar{A}_0^2}+w_7\frac{(\partial\psi)^2}{2a^2}-\alpha\delta\rho_M\notag\\&-(\rho_M+P_M)\frac{(\partial_i v)(\partial^i \chi)}{a^2}-v\delta\dot{\rho}_M-3H(1+c_M^2)v\delta\rho_M-\frac{1}{2}(\rho_M+P_M)\frac{(\partial v)^2}{a^2}-\frac{c_M^2}{2(\rho_M+P_M)}\delta\rho_M^2\Bigg],\\ \label{Lzeta}
               \mathcal{L}_{\zeta}=a^3&\Bigg[3\left(\frac{w_2}{\bar{A}_0}\delta A-w_1\alpha-(\rho_M+P_M)v\right)\dot{\zeta}-\frac{2}{a^2}\left[q_T(\partial_i\chi)+\alpha_3(\partial_i\psi)\right]\partial^i\dot{\zeta}-3q_T\dot{\zeta}^2\notag\\&+\big[2(q_T-2\bar{A}_0\alpha_3)\partial_i\alpha+2\alpha_3(\partial_i\delta A)\big]\frac{\partial^i\zeta}{a^2}+\mathcal{F}_T\frac{(\partial\zeta)^2}{a^2}\Bigg],\\
               \label{LE}\mathcal{L}_{E}=a^3&\Bigg[-2q_T(\partial_i\Ddot{\zeta})-2B_2(\partial_i\dot{\zeta})-w_1(\partial_i\dot{\alpha})-(\dot{w}_1+3Hw_1)\partial_i\alpha+\frac{w_2}{\bar{A}_0}(\partial_i\dot{\delta A})-B_5(\partial_i\delta A)\notag\\&-(\rho_M+P_M)\left[\partial_i\dot{v}-3Hc_M^2(\partial_i v)\right]\Bigg]\partial^i E.
           \end{align}
           \end{widetext}
           The Lagrangian $\mathcal{L}^\mathrm{flat}$ consists of all terms arising in the flat gauge ($\zeta=0,\,E=0$), while $\mathcal{L}_\zeta$ and $\mathcal{L}_E$ capture the contributions from the metric perturbations $\zeta$ and $E$, respectively.\footnote{We remark that we are not choosing any gauge and using $\mathcal{L}^\mathrm{flat}$ is simply a convenient way of denoting this part of the Lagrangian. Moreover, the flat gauge is ill-suited to describe the bounce since $\zeta=0$, which causes apparent strong coupling in the scalar sector when $H=0$.} The coefficients appearing in the Lagrangians above are given in Appendix \ref{app}.
            Notice that the scalar fields $\alpha, \chi, \delta A, v$ and $E$ in the scalar action \eqref{actionS} do not have kinetic terms, which renders them non-dynamical and hence turns their equations of motion into constraints. Varying the action \eqref{actionS} with respect to $\alpha, \chi, \delta A, v, E$, we get the following equations of motion in Fourier space

            \begin{align}
                \label{eomalpha}
                &\mathcal{E}_\alpha=-\delta\rho_M+2w_4\alpha+w_8\frac{\delta A}{\bar{A}_0}-3w_1\Dot{\zeta}-\frac{k^2}{a^2}[\mathcal{Y}+w_1\chi\notag\\&\hspace{20pt}-2(q_T-2\bar{A}_0\alpha_3)\zeta-a^2 w_1\dot{E}]=0,\\[10pt]
                &\mathcal{E}_\chi=-(\rho_M+P_M)v-w_1\alpha+\frac{w_2}{\bar{A}_0}\delta A-2q_T\Dot{\zeta}=0,\\[10pt]
                &\mathcal{E}_{\delta A}=w_8\alpha+2w_5\frac{\delta A}{\bar{A}_0}+3w_2\Dot{\zeta}+\frac{k^2}{a^2}\left(\frac{1}{2}\mathcal{Y}+w_2\chi\right.\notag\\&\hspace{20pt}\left.+2\bar{A}_0\alpha_3\zeta-\frac{\bar{A}_0 w_6-w_2}{2\bar{A}_0}\psi -a^2 w_2 \dot{E} \right)=0,\\[10pt]
                &\mathcal{E}_{v}=\Dot{\delta\rho}_M+3H(1+c_M^2)\delta\rho_M+3(\rho_M+P_M)\Dot{\zeta}\notag\\&\hspace{20pt}+\frac{k^2}{a^2}(\rho_M+P_M)(v+\chi-a^2\dot{E})=0,\label{eomv}\\[10pt]
                &\mathcal{E}_{E}=2q_T\Ddot{\zeta}+2B_2\dot{\zeta}+w_1\dot{\alpha}+(\dot{w}_1+3Hw_1)\alpha-\frac{w_2}{\bar{A}_0}\dot{\delta A}\notag\\&\hspace{20pt}+B_5\delta A+(\rho_M+P_M)(\dot{v}-3Hc_M^2v)=0,
                \label{eomE}
            \end{align}
            where we have abbreviated
            \begin{equation}
                \mathcal{Y}\equiv-\frac{w_3}{\bar{A}_0}(\dot{\psi}-2\bar{A}_0\alpha+\delta A).
            \end{equation}
            We can solve equations \eqref{eomalpha}-\eqref{eomv} for the non-dynamical variables $\alpha, \chi, \delta A$ and $v$. After substituting their solutions into the second-order scalar action \eqref{actionS}, the resulting action does not depend on the non-dynamical scalar $E$ anymore but only on the perturbations $\zeta, \psi$ and $\delta\rho$, meaning that $E$ gets integrated out simultaneously with the former four non-dynamical scalars. This can also be seen from the fact that the equation of motion $\mathcal{E}_E$ \eqref{eomE} becomes trivially satisfied after plugging in the explicit expressions for the non-dynamical fields $\alpha, \chi, \delta A$ and $v$. In the next step, we introduce the gauge-invariant curvature perturbation
            \begin{equation}
                \mathcal{R}_\psi\equiv \zeta-\frac{H}{\bar{A}_0}\psi,
            \end{equation}
            and express the scalar perturbation $\psi$ and its derivatives in terms of $\mathcal{R}_\psi$ in the scalar action. Upon taking the small-scale limit $k^2/a(t)^2\gg H$, we observe that the kinetic term proportional to $\dot{\zeta}^2$ effectively drops out of the action. Given that no mixed derivative terms of the form $\dot{\zeta}\dot{\mathcal{R}}_\psi$ appear, we conclude that $\zeta$ is another auxiliary field with a constraint equation of motion. After integrating terms with $\dot{\zeta}$ by parts, the resulting action contains terms proportional to $\zeta^2$ as well as products of $\zeta$ with dynamical fields, e.g. $\zeta \mathcal{R}_\psi$. Computing the EOM for $\zeta$ from this action allows us to safely integrate out this last non-dynamical variable and leaves us with the dynamical small-scale action 
            \begin{equation}
                \label{Sdyn}
                S_S^{(2)}\simeq\int dt\,d^3x\,a^3\left[\Dot{\Vec{\mathcal{X}}}^T\mathbf{K}\Dot{\Vec{\mathcal{X}}}-\frac{k^2}{a^2}\Vec{\mathcal{X}}^T\mathbf{F}\Vec{\mathcal{X}}\right],
            \end{equation}
            where the only two dynamical scalar modes are encoded in the vector
            \begin{equation}
                \Vec{\mathcal{\chi}}^T=(\mathcal{R}_\psi,\delta\rho_M/k).
            \end{equation}
            $\mathbf{K}$ and $\mathbf{F}$ denote $2\times 2$ matrices that incorporate the kinetic and gradient coefficients of the dynamical perturbations. Notice that no mass terms $\sim \Vec{\mathcal{\chi}}^T\mathbf{M}\Vec{\mathcal{\chi}}$ arise in the action \eqref{Sdyn} because they are dominated by the gradient terms at small scales and can hence be safely neglected within this approximation. For the kinetic and gradient terms, we find the following non-vanishing matrix elements that are relevant for the no-go theorem
            \begin{align}
            \label{QS1}
                &K_{11}\equiv\mathcal{Q}_S=\frac{q_Tq_S}{(w_1-2w_2)^2},\\[10pt]
                \label{qM}
                &K_{22}\equiv\mathcal{Q}_M=\frac{a^2}{2(\rho_M+P_M)},\\[10pt]
                &F_{11}\equiv\mathcal{F}_S=\frac{q_Tq_Sc_S^2}{(w_1-2w_2)^2}\\[10pt]
                &F_{22}\equiv\mathcal{F}_M=\frac{a^2 c_M^2}{2(\rho_M+P_M)},
                \label{FM}
            \end{align}
            where 
            \begin{align}
                &q_S=3w_1^2+4q_Tw_4.
            \end{align}
            Note that the matter sound speed $c_M^2$ is expressed in equation \eqref{cM}. The gauge-invariant scalar sound speed $c_S^2$ was already computed in \cite{Heisenberg:2018wye} and is given by
            \begin{equation}
                c_S^2=\frac{1}{\mathcal{Q}_S}\left(-\mathcal{F}_T+\frac{2E_3^2}{q_V}-\frac{2q_T^2(\rho_M+P_M)}{(w_1-2w_2)^2}+\frac{1}{a}\frac{d}{dt}(aE_2)\right),
            \end{equation}
            where we used the following abbreviations
            \begin{align}
                &E_1\equiv\frac{w_6}{4\bar{A}_0}-\frac{w_1w_2}{4\bar{A}_0^2(w_1-2w_2)},\\[10pt]
                \label{E2}
                &E_2\equiv-\frac{2q_T^2}{w_1-2w_2},\\[10pt]
                &E_3\equiv-\alpha_3-\frac{q_Tw_2}{\bar{A}_0(w_1-2w_2)}.
            \end{align}
            To avoid ghost and gradient instabilities in the (metric) scalar and matter sector, we demand $\mathcal{Q}_S>0, \mathcal{F}_S\geq0, \mathcal{Q}_M>0$ and $\mathcal{F}_M\geq0$. 

        \subsection{Advantage of bouncing solutions with generalized Proca}
        \label{advantage}
            In the previous subsections we have  the perturbative treatment of Generalized Proca theory in a FLRW setting, in the following we briefly review why realizing a bouncing solution with generalized Proca that is stable throughout the entire cosmic evolution appears to present an advantage \cite{Heisenberg:2018wye} compared to bounces with Horndeski theory, which are affected by the no-go theorem \cite{Kobayashi:2016xpl}. This argument motivated the exploration of non-singular bouncing models in generalized Proca theory.

            In Horndeski theory, the following relation holds
            \begin{equation}
                \label{eq}
                \frac{1}{a} \frac{d}{d t}\left(a E_2^{(\rm H)}\right)=\mathcal{Q}_S^{(\mathrm{H})} c_S^{2 (\rm H)}+q_T^{(\mathrm{H})} c_T^{2 (\rm H)}+\frac{2 q_T^{2 (\rm H)}\left(\rho_M+P_M\right)}{w_1^{2 (\rm H)}},
            \end{equation}
            where $E_2^{(\rm H)}$ is the quantity corresponding to \eqref{E2} in Horndeski theories, and all the stability coefficients for perturbations have the superscript $^{(\rm H)}$ to indicate that they are calculated for Horndeski theories.

            The terms on the right-hand side of equation \eqref{eq} are all positive to ensure the absence of ghost and Laplacian instabilities, which means that the following inequality holds
            \begin{equation}
                \frac{1}{a} \frac{d \xi}{d t}>q_T^{(\mathrm{H})} c_T^{2 (\rm H)}>0,
                \label{ineq}
            \end{equation}
            where
            \begin{equation}
                \xi \equiv a E_2^{(\rm H)}=-\frac{2 a q_T^{2(\rm H)}}{w_1^{(\rm H)}}.
                \label{xi}
            \end{equation}

            In order to check whether non-singular bounces are stable throughout the entire evolution, we integrate \eqref{ineq} from $t=t_i$ to $t=t_f\left(>t_i\right)$, and obtain
            \begin{equation}
                \xi_f-\xi_i>\int_{t_i}^{t_f} a q_T^{(\mathrm{H})} c_T^{2 (\rm H)} d t>0.
                \label{int}
            \end{equation}

            The limit $q_T^{(\mathrm{H})} c_T^{2 (\rm H)} \rightarrow 0$ corresponds to either $q_T^{(\rm H)} \rightarrow 0$ or $c_T^{2 (\rm H)} \rightarrow 0$. In the first case, we encounter a strong coupling problem of the tensor perturbations, while the second case directly causes Laplacian instabilities in the tensor sector. 

            The term $q_T^{(\mathrm{H})} c_T^{2 (\rm H)}$ does not tend to $0$ in the asymptotic future, therefore the integral in \eqref{int} is a positive growing function of $t_f$, which translates to $\xi_f>0$ for sufficiently large $t_f$. However, the integral also increases in the asymptotic past $\left(t_i \rightarrow-\infty\right)$, so the condition $\xi_i<0$ is also required. This necessarily means that the function $\xi$ crosses $0$ at some time between $-\infty<t<\infty$, thus $a=0$ in \eqref{xi}. However, this is a contradiction with the fact that we are looking for a consistent non-singular bouncing solution, where $a>0$ by definition throughout the whole cosmological evolution. This argument is proven in the unitary gauge, but also holds in other gauges where the perturbations are well-defined at the bounce, such as the Newtonian gauge \cite{Heisenberg:2018wye}. 

            The situation changes in generalized Proca, as can be seen in the uniform vector gauge characterised by initially setting $\psi=0,\, E=0$.
            Equation \eqref{eq} for generalized Proca theory becomes 
            \begin{equation}
                \frac{1}{a} \frac{d}{d t}\left(a E_2\right)=\mathcal{Q}_S c_S^2+q_T c_T^2+\frac{2 q_T^2\left(\rho_M+P_M\right)}{\left(2 H q_T+w_2\right)^2}-\frac{2 E_3^2}{q_V},
                \label{eqproca}
            \end{equation}
            where the coefficients $E_2,\,E_3$ are given explicitly in the Appendix \ref{app}.
            Compared to \eqref{eq} in Horndeski theory, there is now an additional term $-2 E_3^2 / q_V$, which arises purely from the existence of intrinsic vector modes. Since $q_V>0$ is required for the absence of vector ghosts, the term $-2 E_3^2 / q_V$ needs to be negative. Thus, unlike in Horndeski theory, the right hand side of \eqref{eqproca} is no longer bounded from below by the term $q_T c_T^2$.

            Integrating the expression above from $t=t_i$ to $t=t_f$ yields
            \begin{equation}
                \xi_f-\xi_i=\int_{t_i}^{t_f} a\left[\mathcal{Q}_S c_S^2+q_T c_T^2+\frac{2 q_T^2\left(\rho_M+P_M\right)}{\left(2 H q_T+w_2\right)^2}-\frac{2 E_3^2}{q_V}\right] d t,
            \end{equation}
            where now
            \begin{equation}
                \xi\equiv a E_2=2 a q_T^2 /\left(2 H q_T+w_2\right).
            \end{equation}

            If the term $-2 E_3^2 / q_V$ dominates over the other terms in the asymptotic past $\left(t_i \rightarrow-\infty\right)$, then the integral goes to $-\infty$ and hence $\xi_i>0$. If the term $-2 E_3^2 / q_V$ is subdominant compared to $q_T c_T^2(>0)$ in the asymptotic future $\left(t_f \rightarrow \infty\right)$, the integral grows toward $\infty$ and hence $\xi_f>0$. In this case, it is possible to keep $\xi>0$ throughout the entire cosmological evolution. Therefore, generalized Proca theory holds an advantage over Horndeski theory to construct non-singular bouncing solutions, which is the motivation for the present study.

    
    \section{No-go theorem for a bounce in generalized Proca theory and beyond}
    \label{nogo}
    \subsection{Stability constraints}
            We start by collecting all the stability conditions on the tensor, vector, scalar and matter sector established in Section \ref{pert}
            \begin{align}
            \label{stability}
                &q_T>0,\hspace{10pt}\mathcal{F}_T\geq0,\hspace{10pt}q_V>0,\hspace{10pt}\mathcal{F}_V\geq0,\notag\\[10pt]
                &q_S>0,\hspace{10pt}\mathcal{F}_S\geq0,\hspace{10pt}\rho_M+P_M\geq 0,\hspace{10pt}c_M^2\geq0.
            \end{align}
            If a stable bounce is to be realized, all the conditions above must be fulfilled not only at the critical moment of the bounce  $t=t_B$, but at all times in the cosmological time evolution.

        \subsection{No-go theorem}
        \label{symmetry}
        In this section, we show that non-singular bouncing cosmologies in generalized Proca theory unavoidably run into matter, scalar or tensor instabilities. These results amount to a \textit{no-go theorem} for such bounces in generalized Proca theory and beyond. 

        We start from the fact that avoiding instabilities in the matter sector requires $\mathcal{Q}_M>0$. According to \eqref{qM}, this is satisfied if and only if $\rho_M+P_M\geq 0$. In the following, we show that both $\rho_M+P_M>0$ and $\rho_M+P_M=0$ separately lead to instabilities, either in the tensor or the scalar sector.
        
        Let us first consider the case of $\rho_M+P_M>0$. Combining the background equations \eqref{rho}-\eqref{p}, we find
        \begin{align}
            \label{rho+p1}
            &\rho_M+P_M=-4G_4\Dot{H}+4\bar{A}_0^2G_{4,X}\Dot{H}+2\bar{A}_0^3HG_{5,X}\Dot{H}\notag\\&-\bar{A}_0^2G_{3,X}\dot{\bar{A}}_0+4\bar{A}_0HG_{4,X}\dot{\bar{A}}_0+4\bar{A}_0^3HG_{4,XX}\dot{\bar{A}}_0\notag\\&+3\bar{A}_0^2H^2G_{5,X}\dot{\bar{A}}_0+\bar{A}_0^4H^2G_{5,XX}\dot{\bar{A}}_0\notag\\&=-\left(2q_T\Dot{H}+w_2\frac{\dot{\bar{A}}_0}{\bar{A}_0}\right).
        \end{align}

        The coefficients $q_T$ and $w_2$ are given in \eqref{qT} and \eqref{w2}, respectively. From their definitions, we see that they only depend on $\bar{A}_0$ and $H$, with $\bar{A}_0$ being fully determined by $H$ as well through the background EOM \eqref{A0eq}. 
        Taking the derivative of the equation of motion for $\bar{A}_0$ found in \eqref{A0eq} with respect to $t$, we obtain
        \begin{widetext}
        \begin{align}
        \label{dtA0eq21}
            \frac{d}{dt}\left(\frac{\mathcal{E}_{\bar{A}_0}}{\bar{A}_0}\right)&=\bar{A}_0G_{2,XX}\dot{\bar{A}}_0-3HG_{3,X}\dot{\bar{A}}_0-3\bar{A}_0G_{3,X}\Dot{H}-3\bar{A}_0^2HG_{3,XX}\dot{\bar{A}}_0+12HG_{4,X}\Dot{H}+18\bar{A}_0H^2G_{4,XX}\dot{\bar{A}}_0\notag\\&\hspace{10pt}+12\bar{A}_0^2HG_{4,XX}\Dot{H}+6\bar{A}_0^3H^2G_{4,XXX}\dot{\bar{A}}_0+3H^3G_{5,X}\dot{\bar{A}}_0+9\bar{A}_0H^2G_{5,X}\Dot{H}+6\bar{A}_0^2H^3G_{5,XX}\dot{\bar{A}}_0\notag\\&\hspace{10pt}+3\bar{A}_0^3H^2G_{5,XX}\Dot{H}+\bar{A}_0^4H^3G_{5,XXX}\dot{\bar{A}}_0\notag\\[10pt]&=\frac{2w_5}{\bar{A}_0^3}\dot{\bar{A}}_0-\frac{3w_2}{\bar{A}_0^2}\Dot{H}=0,
        \end{align}
        \end{widetext}
        where we have used the coefficients $w_2$ and $w_5$ given in \eqref{w2} and \eqref{w5}, respectively. Note that $\bar{A}_0\neq 0$, since $\bar{A}_0=0$ corresponds to the trivial solution of the background equation \eqref{A0eq}. 
        Combining \eqref{dtA0eq21} and \eqref{rho+p1}, we find 
        \begin{equation}
            \rho_M+P_M=-\Big(2q_T+\frac{3w_2^2}{2w_5}\Big)\dot{H}.
        \end{equation}
        We now specify to the moment of the bounce: since $\dot{H}(t_B)>0$, equation \eqref{rho+p1} tells us that $\rho_M+P_M(t_B)$ is positive if
        \begin{equation}
            \label{qTineq}
            0<2q_T(t_B)<-\frac{3w_2^2(t_B)}{2w_5(t_B)},
        \end{equation}
        where the first inequality is needed for a stable tensor sector at the bounce. Next, we explicitly evaluate $w_2$ given in \eqref{w2} at $t_B$ by setting $H(t_B)=0$, which yields 
        \begin{equation}
            \label{w2B}
            w_2(t_B)=\bar{A}_0^3(t_B) G_{3,X}(X_B),
        \end{equation}
        with $X_B=X(t_B)$ indicating the value of the vector field norm at the bounce.
         A complementary constraint comes from finding the derivative with respect to $H$ of the equation of motion for $\bar{A}_0$ \eqref{A0eq}, which, at the moment of the bounce $t_{B}$, reads
        \begin{equation}
        \label{dHA0eq}
            \frac{d}{dH}\left(\frac{\mathcal{E}_{\bar{A}_0}}{\bar{A}_0}\right)\Bigg |_{H(t_B)=0}=-3\bar{A}_0(t_B)G_{3,X}(X_B)=0,
        \end{equation}
        where we have again discarded the trivial solution branch $\bar{A}_0=0$. Combining the above equation \eqref{dHA0eq} with the expression for $w_2(t_B)$ in \eqref{w2B}, we see that if $\bar{A}_0(t_B)=0$, it directly follows that $w_2(t_B)=0$. Alternatively, if $A_0(t_B)\neq0$, equation \eqref{dHA0eq} then enforces $G_{3,X}(X_B)=0$, which still leads to $w_2(t_B)=0$. Both statements combined translate to $q_T(t_B)=0$ via the inequality \eqref{qTineq} and therefore cause strong coupling in the tensor sector at the moment of the bounce, which prohibits a stable bounce.
        The crucial issue at the heart of the reasoning above (and the difference with respect to bounces driven by Horndeski theories) is that $\bar{A}_0$ is not a dynamical degree of freedom, but plays the role of an auxiliary field. Its dynamics stems solely from the Hubble parameter $H(t)$, as can be observed from the background equation of motion \eqref{A0eq}.

        Let us now consider the case of $\rho_M+P_M=0$: we show that one incurs instead into a strong coupling problem in the scalar sector. We rewrite the coefficient of the scalar kinetic term as
        \begin{equation}
        \label{QS2}
            \mathcal{Q}_S=q_T\frac{3w_2^2+4w_5q_T}{(w_1-2w_2)^2},
        \end{equation}
        using equations \eqref{w2} and \eqref{w5} to replace $w_1$ and $w_4$ in the original expression \eqref{QS1}. To prevent scalar strong coupling, we must have $\mathcal{Q}_S>0$. Looking at equation \eqref{dtA0eq21} in the bouncing phase where $\dot{H}>0$, we observe the following: if $w_5=0$, equation \eqref{dtA0eq21} implies that $w_2=0$ and hence $\mathcal{Q}_S=0$, which is not allowed. Thus, it follows that $w_5\neq 0$. If $w_2=0$, setting equation \eqref{rho+p1} to zero implies $q_T=0$ and we hence run into tensor strong coupling. Therefore, we conclude that $w_2\neq 0$ as well. Now, we can recast equation \eqref{dtA0eq21} as
        \begin{equation}
        \label{eq11}
            \frac{\dot{\bar{A}}_0}{\bar{A}_0\Dot{H}}=\frac{3w_2}{2w_5}.
        \end{equation}
        Given $\rho_M+P_M=0$, equation \eqref{rho+p1} yields
        \begin{equation}
        \label{eq21}
            \frac{\dot{\bar{A}}_0}{\bar{A}_0\Dot{H}}=-\frac{2q_T}{w_2}.
        \end{equation}
        Equating the right-hand side of equations \eqref{eq11} and \eqref{eq21}, we arrive at
        \begin{equation}
            3w_2^2=-4w_5q_T,
        \end{equation}
        and hence, it clearly follows from \eqref{QS2} that $\mathcal{Q}_S=0$ and the scalar becomes strongly coupled. Overall, we have shown that under the assumption $\Dot{H}>0$, which is valid during the bounce, demanding $\rho_M+P_M\geq0$ to avoid matter instabilities directly leads to either tensor strong coupling at the moment of the bounce or scalar strong coupling. This concludes our no-go theorem for cosmological bounces mediated by a generalized Proca field.

\subsection{Complementary no-go argument based on background equations}
\label{background}
A strikingly simple complementary argument showing the impossibility of a stable bounce in generalized Proca can be formulated by paying particular attention to the perturbative stability of the matter sector, and specifically the matter sound speed \eqref{cM}. This quantity exhibits the most persistent instability: even when all other coefficients fulfil the conditions \eqref{stability}, $c_M^2$ diverges at the moment of the bounce. Stability is spoiled since positive powers of $c_M$ appear in the second-order action \eqref{Lflat} (and consequently in the equations of motion), causing it to diverge. This persistent instability is again caused by the non-dynamical nature of $\bar{A}_0$ and the peculiar relationship existing between its EOM and the expression at the denominator of $c_M^2$, as we clarify in the following.

From the non-trivial branch of the equation of motion for $\bar{A}_0$ \eqref{A0eq}, one finds
\begin{align}
    G_{2,X}=&3G_{3,X}H\bar{A}_0-6H^2(G_{4,X}+\bar{A}_0^2G_{4,XX})\notag\\
    &-\bar{A}_0H^3(3G_{5,X}+\bar{A}_0^2G_{5,XX}),
\end{align}
which, at the bounce, reads
\begin{equation}
\label{zero}
    G_{2,X}(X_B)=0,
\end{equation}
given that $H(t_B)=0$. Substituting the equation above in the background equation \eqref{rho} and evaluating it at the bounce, we obtain
\begin{equation}
\rho_M(t_B)=G_2(X_B),
\end{equation}
and thus 
\begin{equation}
\label{dotrho}
    \dot{\rho}_M(t_B)=G_{2,X}(X_B)\bar{A}_0(t_B)\dot{\bar{A}}_0(t_B)=0,
\end{equation}
if we take \eqref{zero} into account.
This is relevant since the denominator of $c_M^2$ is precisely $\dot{\rho}_M$ by definition, and since this denominator vanishes at $t_B$, the matter sound speed squared $c_M^2$ always diverges at the moment of the bounce if the background equations are satisfied.

The argument above is generic and valid for any choice of coupling $G_2$, as it only relies on the validity of the background equations. 

\subsection{Robustness of no-go result}
The two no-go results above have been proven without making any gauge choice. The first one in \ref{symmetry} relies on the stability coefficients \eqref{stability}, that have been computed from the gauge-ready form of the second-order action \eqref{actionS}. The second argument in \ref{background} relies purely on the background equations. Therefore, the result of our work is robust to different gauge choices, although the standard in the literature is usually to study the question of bounce stability simply in a particular gauge, where the treatment is simpler. For example, the no-go theorem for Horndeski non-singular bounces \cite{Kobayashi:2016xpl} is proven in the unitary gauge. However, this result has been debated on the basis on considerations made in the Newtonian gauge \cite{Ijjas:2017pei}.\\

        Given the success of beyond Horndeski theory to circumvent the no-go theorem, one might wonder whether beyond generalized Proca theory might provide a similar way out. However, this is not possible: as noted in \cite{Heisenberg:2016eld} and explained in section \ref{proca}, beyond generalized Proca theory is indistinguishable from generalized Proca theory at the background level because the two additional couplings $B_4, B_5$ arising from $\mathcal{L}_4^N, \mathcal{L}_5^N$ do not contribute to the background equations. This means that the non-dynamical function $\bar{A}_0(t)$ depends on time through the cosmological background $H(t)$ only, which lies at the heart of our no-go theorem, and therefore leaves it unchanged for beyond Proca theory as well. Furthermore, the matter action is not affected by the extension to beyond generalized Proca theory, so $q_M$ looks identical compared to the generalized Proca case. As both the tensor and scalar kinetic coefficients $q_T$ and $\mathcal{Q}_S$ take similar forms in beyond generalized Proca theory because the coefficients $q_T$, $w_2$ and $w_5$ do not depend on $B_4, B_5$, demanding $\rho_M+P_M\geq 0$ for stability in the matter sector leads to the same tensor and scalar strong coupling problem. Therefore, the no-go theorem for bounces still holds in beyond generalized Proca theory and cannot be evaded.


    \section{Discussion and conclusions}
        Fundamental shortcomings of inflation, such as the trans-Planckian problem, have spurred the search for alternatives, with one intriguing possibility given by non-singular classical bounces. Bouncing cosmologies require a NEC-violating phase, which cannot be accommodated within GR coupled to conventional matter, but can be realized with modified theories of gravity. In practice, however, cosmological bounces mediated by additional DOF prove difficult to be constructed in a healthy manner, i.e.\ within weakly-coupled theories free of ghost, gradient and Laplacian instabilities. A notable exception is provided by beyond Horndeski theory, for which a number of fully stable bouncing cosmologies have been found \cite{Mironov:2019haz, Cai:2017dyi, Kolevatov:2017voe}. Another possibility worth investigating is generalized Proca theory: it appears to present an advantage over Horndeski theory, which is affected by a no-go theorem. 
        
        However, the present work shows that generalized Proca theory is also plagued by a similar (and even more robust) no-go theorem. This result has been proven following two complementary arguments, one based on perturbative instabilities in the tensor and scalar sector and one relying on the background equations. On the one hand, demanding a stable matter sector during the bounce unavoidably implies strong coupling of either the tensor or scalar DOF. On the other hand, by simply combining the background equations at the bounce, one finds a persistent divergence of the speed of sound for matter perturbations. This divergence is directly related to the temporal component of the background vector field, $\bar{A}_0$, being only an auxiliary field, such that all of its dynamics stems from its dependence on the cosmological background $H(t)$ (the non-dynamical character of this component is required in order to keep three healthy propagating degrees of freedom in the theory, and is therefore essential \cite{Heisenberg:2014rta}). It is quite intuitive that a non-dynamical field is not able to drive the bounce, since its dynamics depends on $H$ only. Examples of non-singular bounces with non-dynamical scalar fields generally rely on other mechanisms to achieve the bounce, such as limiting curvature (see, e.g. \cite{Quintin:2019orx}).
        Crucially, our results have been proven without resorting to any specific gauge choice, but relying on a gauge-ready formulation, which makes them more robust.
        There are notable differences between the no-go theorem for Horndeski theories \cite{Libanov:2016kfc, Kobayashi:2016xpl} and the one found in the present work. The first establishes the impossibility of constructing a non-singular bouncing model with Horndeski theories that is stable \textit{throughout the entire evolution of the universe}, thus including the requirement that gravity is modified only to achieve the bounce and then reduces to GR asymptotically in the past and future. The no-go theorem for non-singular bounces in generalized Proca theory instead states that the bounce \textit{itself} is inevitably unstable, and not that instabilities are encountered at some point in the entire evolution of the universe. Despite the promising general behaviour of generalised Proca described in \ref{advantage}, a stable bounce cannot be realized because of fundamental issues that ultimately trace back to the non-dynamical nature of the temporal component of the vector field. Beyond generalized Proca theory cannot remedy the situation as it is indistinguishable from ordinary generalized Proca at the background level, where the non-dynamical nature of the temporal component of the vector field is established.

        In the present work, we restricted our analysis to linear perturbations only, which is the first step in testing whether a classical, non-singular bouncing model is viable. However, this is not sufficient in general: a successful bounce should also incorporate a mechanism to prevent the growth of anisotropies during the contracting phase to ensure that they do not disrupt the bounce. For example, this can be achieved through an \textit{ekpyrotic} phase \cite{Khoury:2001wf}, or a phase with a super stiff equation of state. Preventing the growth of anisotropies is required in order to avoid the onset of the Belinskii-Khalatnikov-Lifshitz (BKL) instability \cite{Belinsky:1970ew} and the chaotic mixmaster behaviour \cite{Misner:1969hg} that a contracting universe incurs in when approaching the singularity. 
        Of course, we have not dealt with such issues in the present work, as the first step of linear stability is already unfulfilled by bounces driven by generalized Proca.
        
        The analysis in this work also includes the contribution of a baryonic matter action \eqref{schutzsorkin} that accounts for the matter present in the universe, but eventually turns out to be one of the sources of instabilities. We remark that bounces based on Horndeski usually do not include a matter contribution because its contribution to the energy budget is thought to remain subdominant with respect to the scalar driving the bounce. Crucially, also the no-go theorem for bounces in Horndeski theory \cite{Kobayashi:2016xpl} neglects the presence of matter. Without the inclusion of matter, some instabilities might have been avoided in the generalized Proca bounce, but we deem our approach to be more realistic. 
        
        There exist more general SVT theories \cite{Heisenberg:2018wye} that feature both a higher number of tunable DOF and different background dynamics than generalized Proca theory; hence, they present the possibility of avoiding our no-go theorem for generalized Proca theory and beyond (for progress in this direction, see e.g. \cite{Heisenberg:2018wye}). 
        Therefore, generalized Proca theory may still play a part in realizing a healthy bounce that leads to the present-day universe if suitably combined beyond Horndeski theory. It would be also interesting to investigate whether the Proca Nuevo theories \cite{deRham:2021efp} can avoid the reported difficulties here. The investigation of these issues is left for future work.

        \begin{acknowledgments} 
            The authors are grateful to Jerome Quintin and Hannes Heisler for thoughtful feedback on this work. S.G. and N.N. also thank Jean-Luc Lehners and Patrick Peter for useful discussions. L.H. is supported by funding from the European Research Council (ERC) under the European Unions Horizon 2020 research and innovation programme grant agreement No 801781. L.H. further acknowledges support from the Deutsche Forschungsgemeinschaft (DFG, German Research Foundation) under Germany’s Excellence Strategy EXC 2181/1 - 390900948 (the Heidelberg STRUCTURES Excellence Cluster). 
        \end{acknowledgments}
\appendix\section{Coefficients in scalar second-order action}
\label{app}
In this Appendix, we provide the cumbersome expressions of some coefficients involved in the second-order action for the scalar sector \eqref{Lflat}-\eqref{LE}, namely:
\begin{align}
                &w_1=\bar{A}_0^3G_{3,X}-4H(G_4+\bar{A}_0^4G_{4,XX})\notag\\&\hspace{23pt}-H^2\bar{A}_0^3(G_{5,X}+\bar{A}_0^2G_{5,XX}),\\[10pt]
                \label{w2}
                &w_2=w_1+2Hq_T,\\[10pt]
                &w_3=-2\bar{A}_0^2q_V,\\
                &w_4=\frac{1}{2}\bar{A}_0^4G_{2,XX}+\frac{3}{2}H\bar{A}_0^3(G_{3,X}-\bar{A}_0^2G_{3,XX})\notag\\&\hspace{23pt}-3H^2(2G_4+2\bar{A}_0^2G_{4,X}+\bar{A}_0^4G_{4,XX}-\bar{A}_0^6G_{4,XXX})\notag\\&\hspace{23pt}-\frac{1}{2}H^3\bar{A}_0^3(9G_{5,X}-\bar{A}_0^4G_{5,XXX}),\\
                \label{w5}
                &w_5=w_4-\frac{3}{2}H(w_1+w_2),\\[10pt]
                &w_6=-\bar{A}_0[\bar{A}_0G_{3,X}+4H(G_{4,X}-\bar{A}_0^2G_{4,XX})\notag\\&\hspace{23pt}+\bar{A}_0H^2(G_{5,X}-\bar{A}_0^2G_{5,XX})],\\[10pt]
                &w_7=[G_{3,X}-4\bar{A}_0HG_{4,XX}-H^2(G_{5,X}+\bar{A}_0^2G_{5,XX})]\Dot{\bar{A}}_0\notag\\&\hspace{23pt}-2\Dot{H}(2G_{4,X}+\bar{A}_0HG_{5,X}),\\[10pt]
                &w_8=3Hw_1-2w_4,\\
                &\alpha_3=\frac{1}{4H}\left(w_6+\frac{w_2}{\bar{A}_0}\right),\\[10pt]
                &B_2=\dot{q}_T+3Hq_T,\\[10pt]
                &B_5=-\frac{1}{\bar{A}_0}\big[\dot{w}_2+3Hw_2+\dot{\bar{A}}_0(w_6-4H\alpha_3)\big].
            \end{align}
    \bibliographystyle{apsrev4-2}
    \bibliography{main.bib}

\end{document}